\documentclass[12pt, amsmath, amssymb]{revtex4}

\usepackage{amsmath,amssymb,amsthm,amsfonts}
\usepackage{enumerate}
\usepackage{epsfig}
\usepackage{graphicx}
\usepackage{pgf}

\usepackage{natbib}
\bibpunct{}{}{,}{s}{,}{,}
\makeatletter
\DeclareRobustCommand\onlinecite{\@onlinecite}
\def\@onlinecite#1{\begingroup\let\@cite\NAT@citenum\citealp{#1}\endgroup}
\makeatother

\newtheorem{theorem}{Theorem}[section]
\newtheorem*{theorem*}{Theorem}

\newtheorem{lemma}[theorem]{Lemma}

\newtheorem*{corollary*}{Corollary}

\newtheorem*{conjecture*}{Conjecture}

\theoremstyle{remark}

\newtheorem*{remark}{Remark}

\theoremstyle{definition}
\newtheorem{definition}[theorem]{Definition}

\newtheorem{algorithm}{Algorithm}

\newcommand{\Z}{\mathbb{Z}}

\begin{document}

\title{An accurate and efficient tau-leaping procedure for the simulation
  of chemical reaction systems}

\author{David F. Anderson}
\email{anderson@math.wisc.edu}
\homepage{www.math.wisc.edu/~anderson}
\affiliation{Department of Mathematics, University of
  Wisconsin-Madison, Madison, Wi 53706}

\date{\today}

%\doublespacing
\begin{abstract}

  By explicitly representing the reaction times of discrete chemical
  systems as the firing times of independent, unit rate Poisson
  processes we develop a new adaptive tau-leaping procedure.  The
  procedure developed is novel in that accuracy is guaranteed by
  performing post-leap checks.  Because the representation we use
  separates the randomness of the model from the state of the system,
  we are able to perform the post-leap checks in such a way that the
  statistics of the sample paths generated will not be skewed by the
  rejections of leaps.  Further, since any leap condition is ensured
  with a probability of one, the simulation method naturally avoids
  negative population values.

\end{abstract}

\maketitle
\section{Introduction}

The procedure developed in this paper is a tau-leaping method for
simulating the evolution of discrete stochastic chemical systems.  The
novelty of the procedure is that a post-leap check is performed after
each step in order to guarantee accuracy.  Post-leap checks have been
avoided in the past because of the worry that rejecting leaps will
skew the statistics of the sample paths.  This problem is bypassed in
our method by storing all the information gained during each leap for
future use.  By performing a post-leap check to ensure accuracy, the
method developed in this paper naturally avoids negative population
values without the need for any extra effort in either a programming
or numerical sense.

Consider a chemically reacting system consisting of $N \ge 1$ chemical
species, $\{X_1,\dots,X_N\}$, undergoing $M \ge 1$ chemical reactions,
each of which is equipped with a propensity function (or intensity
function in the mathematics literature), $a_k(X(t))$, which is a
function of the state of the system at time $t$, $X(t) \in \Z_{\ge
  0}^N$.  Let $\nu_k, \nu_k' \in \Z^N_{\ge 0}$ be the vectors
representing the number of molecules of each species consumed and
created in the $k$th reaction, respectively.  If $R_k(t)$ is the
number of times that the $k$th reaction has taken place up to time
$t$, then the state of the system at time $t$ is given by
\begin{equation*}
  X(t) = X(0) + \sum_{k=1}^M R_k(t)(\nu_k' - \nu_k).
\end{equation*}

The fundamental assumption of stochastic chemical kinetics states that
the probability that reaction $k$ takes place in the infinitesimal
amount of time $[t,t + \Delta t)$ is given by $a_k(X(t))\Delta t +
O(\Delta t^2)$.\cite{Gill76, Gill77} That is, $P(R_k(t + \Delta t) -
R_k(t) = 1 \ | \ X(s), s \le t) = a_k(X(t))\Delta t + O(\Delta t^2)$.
For each $k \le M$, let $Y_k(\cdot)$ be an independent, unit rate
Poisson process.  That is, for any $T>0$, and small $\Delta T > 0$,
$P(Y_k(T + \Delta T) - Y_k(T) = 1) = \Delta T + O(\Delta T ^2)$, for
$j \ne k$.  Because the propensity function of reaction $k$ is
$a_k(X(t))$ until the next reaction takes place,
\begin{align*}
  &P\left(Y_k\left(\int_0^{t+\Delta t} a_k(X(s))ds\right) -
    Y_k\left(\int_0^{t} a_k(X(s))ds\right) = 1 \ | \ X(s), s \le
    t\right) \hspace{.8in} \phantom{.}\\ 
  & \hspace{2in} = a_k(X(t))\Delta t + O(\Delta t^2),
\end{align*}
and we see that $R_k(t)$ can be written as
\begin{equation}
  R_k(t) = Y_k \left(\int_0^t a_k(X(s))ds \right).
  \label{eq:Rk}
\end{equation}
The state of the system at time $t$ can therefore be represented as
the solution to the following stochastic equation
\begin{equation}
  X(t) = X(0) + \sum_{k=1}^M Y_k \left(\int_0^t a_k(X(s))ds \right)
  (\nu_k' - \nu_k).
  \label{eq:state}
\end{equation}
We note that even though the processes $Y_k$ are independent, the
terms $Y_k \left(\int_0^t a_k(X(s))ds \right)$ are dependent because
they depend upon the process $X(s)$, for $s \le
t$.\cite{Andersson2000} Equation \eqref{eq:state} is typically called
a random time change representation in the Mathematics literature.
\cite{Anderson2007a, Ball06, Kurtz78, Kurtz86}

Note that there are $M+1$ distinct time frames in \eqref{eq:state}.
The first time frame is the actual, or absolute time, $t$.  However,
each Poisson process $Y_k$ brings its own ``internal'' time frame.
Equation \eqref{eq:Rk} shows that at absolute time $t$, the amount of
``internal time'' that has passed for the process $Y_k$ is $T_k(t) =
\int_0^t a_k(X(s)) ds$. This observation leads to the following
definition.
\begin{definition}
  For each $k \le M$, $T_k(t) = \int_0^t a_k(X(s)) ds$ is the {\em
    internal time} of the Poisson process $Y_k$ at absolute time $t$.
\end{definition}
\noindent We note that the values $T_k(t)$ defined above do not
actually have units of time.  In fact, they are unit-less.  However,
they serve the purpose of allowing us to know where we are on the
``time frames'' of the Poisson processes $Y_k$ at absolute time $t$.
\begin{remark} It is important to recognize that for any $T_2 \ge T_1
  \ge T_k(t)$, the increment $Y_k(T_2) - Y_k(T_1)$ is independent from
  the state of the system, $X(t)$.  This independence follows from the
  usual properties of Poisson processes and will eventually allow us
  to reject leaps without adding any bias to the system.
\end{remark}

The outline of the paper is as follows.  In Section
\ref{sec:background} we will briefly introduce several exact
simulation methods for chemical systems and a widely used approximate
method known as tau-leaping.  While all of the methods presented in
Section \ref{sec:background} are well known, we will consider each
through the perspective of equations \eqref{eq:Rk} and
\eqref{eq:state}, which we believe lends insight.  In Section
\ref{sec:newtau} we present our new adaptive tau-leaping procedure.
In Section \ref{sec:dimer} we compare the efficiency of our new
algorithm to the current adaptive tau-leaping procedures on a model of
a decaying dimer.

\section{Background}
\label{sec:background}

\subsection{Exact simulation methods}
\label{sec:exact}

In order to generate sample paths for a given system all exact
simulation methods attempt to answer each of the following two
questions at a given moment of time, $t$:
\begin{enumerate}
\item[1a.] When does the next reaction take place?
\item[2a.] Which reaction takes place at that future time?
\end{enumerate}
By answering both questions repeatedly, a sample path is constructed.
We see from equation \eqref{eq:Rk} that the above questions are
equivalent to:
\begin{enumerate}
\item[1b.] What will be the absolute time of the next firing of the
  processes $Y_k(\int_0^t a_k(X(s))ds)$?
\item[2b.] Which process will fire at that time?
\end{enumerate}

Note that neither the state of the system, $X(t)$, nor the propensity
functions, $a_k(X(t))$, change between reactions.  Therefore, assuming
that no other reaction fires first, the next firing time of the
process $Y_k(\int_0^t a_k(X(s))ds)$ will be exponentially distributed
with parameter $a_k(X(t))$.  Of course, the logic used in the previous
sentence is only valid up until the time of the first firing of the
various processes, for at that time the state of the system, and hence
the propensity functions, will change.  Therefore, we may only
conclude that the next firing time of the Processes $Y_k(\int_0^t
a_k(X(s))ds)$ will occur at the minimum time of the exponentially
distributed random variables, and the reaction that takes place is
simply the one associated with the realized minimum value.  Repeated
application of the above idea is the First Reaction
Method.\cite{Gill76} The Next Reaction Method \cite{Gibson2000} and
modified Next Reaction Method \cite{Anderson2007a} use the same
principles as the First Reaction Method except that by efficient use
of information, both need to only generate one exponential random
variable per iteration as opposed to the $M$ needed in the First
Reaction Method.  See Ref.  \onlinecite{Anderson2007a} for full
details about how the Next Reaction Method and modified Next Reaction
Method achieve this efficiency.  The Gillespie Algorithm,\cite{Gill76,
  Gill77} or Stochastic Simulation Algorithm (SSA), answers the first
question by using the fact that the minimum of $M$ exponentially
distributed random variables with parameters $a_k$ is exponentially
distributed with parameter $\sum_{k=1}^M a_k$.  To answer the second
question the Gillespie Algorithm uses the fact that the probability
that the $j$th exponential random variable achieves the minimum is
$a_j/\sum_{k=1}^M a_k$.  Therefore, for every iteration of the
Gillespie Algorithm one random number is needed to find when the next
reaction occurs, and one random number is needed to determine which
reaction occurs at that later time.

We note that the algorithms described above are considered exact
simulation methods because they generate {\em statistically} exact
sample paths for the system \eqref{eq:state}.  Typically one wishes to
use such methods to generate many sample paths in order to approximate
the underlying probability distributions of the system of
interest.\cite{Arkin1998, McAdams1997, Ozbudak2002, Petzold2005}
However, there are instances when the exactness of the methods make
them ineffectual and approximate techniques are needed.

\subsection{An approximate method: tau-leaping}
\label{sec:approx}

Because they simulate every reaction that takes place, statistically
exact methods are slow for systems in which many reactions take place
over short amounts of time.  As the algorithms described in this paper
are typically used for Monte Carlo simulations in which thousands,
tens of thousands, or even hundreds of thousands of sample paths are
needed to get an accurate picture of the underlying probability
distributions, it is clear why simulation speed is critical.
Therefore, approximate techniques have been developed that will
generate sample paths significantly faster than the exact methods and
will do so with an acceptable amount of error.  One such method is
tau-leaping. \cite{Gill2001}

Consider equations \eqref{eq:Rk} and \eqref{eq:state}.  We make the
observation that there are two natural places where we can approximate
the system: the Poisson processes, $Y_k$, and the propensity
functions, $a_k$.  In standard tau-leaping, only the propensity
functions are approximated.  More specifically, if one assumes that
$a_k(X(t))$ is relatively constant in the time interval $[t,t + \tau)$
(this assumption is typically called the {\em leap condition}), then,
conditioned on $X(s)$ for $s \le t$, the number of times the $k$th
reaction fires in the time interval $[t,t+\tau)$ can be approximated by
\begin{align}
  \begin{split}
    \hbox{Number of firings} &= R_k(t + \tau) - R_k(t)\\
    &= Y_k\left(\int_{0}^{t + \tau} a_k(x(s))ds \right) -
    Y_k\left(\int_{0}^{t} a_k(x(s))ds
    \right)\\
    &\approx Y_k\left( a_k(x(t))\tau + \int_0^t a_k(x(s))ds \right) -
    Y_k\left(\int_0^t a_k(x(s))ds\right).
  \end{split}
  \label{eq:approx}
\end{align}
Using that each $Y_k$ is a unit rate Poisson process then gives us
that, conditioned on $X(s)$ for $s \le t$,
\begin{equation}
  Y_k\left( a_k(x(t))\tau + \int_0^t a_k(x(s))ds \right) -
  Y_k\left(\int_0^t a_k(x(s))ds\right) \overset{d}=
  \hbox{Poisson}(a_k(x(t))\tau).
  \label{eq:numfirings}
\end{equation}
Therefore, we use Poisson random variables to approximate how many
times each reaction has fired from time $t$ to $t + \tau$, and we
update the system via
\begin{equation}
  x(t + \tau) = x(t) + \sum_{k = 1}^M N_k(\nu_k' - \nu_k),
  \label{eq:update}
\end{equation}
where $N_k$ is a Poisson random variable with parameter
$a_k(x(t))\tau$.  We note that based upon the approximation used in
\eqref{eq:approx}, tau-leaping is similar to an Euler method.

The subtlety of tau-leaping is in selecting a $\tau$ before each step
so that the leap condition holds over the time interval $[t, t +
\tau)$.  A typical way to make this explicit is to search for a $\tau$
so that for some small $\epsilon > 0$
\begin{equation}
  \left| a_k(X(t + \tau)) -
    a_k(X(t))\right| \le \max\{\epsilon a_k(X(t)),c_k\},
  \label{leapCond1}
\end{equation}
where $c_k$ is the rate constant for reaction $k$, which is the
smallest amount that a propensity function can change.  The question
now becomes how to go about selecting the largest $\tau$ for which we will be
reasonably sure that the condition \eqref{leapCond1} will be
satisfied.

The tau-leaping method proposed by Cao et al. \cite{Cao2006} chooses
$\tau$ before each step to be the largest value for which both the
estimated mean and the estimated standard deviation of the random
variable on the left side of equation \eqref{leapCond1} satisfies that
condition.  It is shown in Ref. \onlinecite{Cao2006} that a
computationally efficient way to do this is to compute the $2N$
quantities
\begin{align*}
  \hat \mu_i(X(t)) = \sum_{j = 1}^M (\nu_j' - \nu_j)_i a_j(X(t)), \ \ \ i =
  1,\dots,  N\\
  \hat \sigma^2_i(X(t)) = \sum_{j=1}^M (\nu_j' - \nu_j)_i^2 a_j(X(t)), \ \ \
  i = 1,\dots,N,
\end{align*}
and then take $\tau$ to be the value given by
\begin{equation}
  \tau = \min_{i \in [1,N]}\left\{\frac{\max\{\epsilon
      X_i(t)/g_i,1\}}{|\hat \mu_i(X(t))|},\frac{(\max\{\epsilon
      X_i(t)/g_i,1\})^2}{\hat \sigma^2_i(X(t))}\right\},
  \label{eq:tau2}
\end{equation}
where $g_i$ for each species $X_i$ is a simple prescribed function of
$X_i(t)$ whose form is fixed at the beginning of the simulation and is
given in Appendix \ref{app:g}.  In computing $\tau$ with the above
method, the leap condition that is actually being satisfied is
\begin{equation}
  |X_i(t + \tau) - X_i(t)| \le \max\{\epsilon X_i(t)/g_i,1\},
  \label{leapCond2}
\end{equation}
which then approximately satisfies the leap condition
\eqref{leapCond1}.  See Ref. \onlinecite{Cao2006} for full details.
The tau-leaping algorithm presented below chooses tau by a pre-leap
computation using equation \eqref{eq:tau2}.

\begin{algorithm} (Cao et al.\cite{Cao2006} pre-leap computation
  tau-leaping)
  \begin{enumerate}
    \setlength{\itemsep}{1pt}
    \setlength{\parskip}{0pt}
    \setlength{\parsep}{0pt}
  \item Initialize.  Set the initial number of molecules of each
    species and set $t = 0$.
  \item Calculate the propensity function, $a_k$, for each reaction.
    \label{taustep}
  \item Calculate $\tau$ according to equation \eqref{eq:tau2}.
    \label{othertaustep}
  \item For each $k \le M$, let $N_k = \hbox{Poisson}(a_k\tau)$.
    \label{step1}
  \item Set $x = x + \sum_{k = 1}^M N_k(\nu_k' - \nu_k)$ and $t = t +
    \tau$.
    \label{step2}
  \item Return to step \ref{taustep} or quit.
  \end{enumerate}
  \label{alg:oldtau}
\end{algorithm}

There are two technical features of standard tau-leaping that remain
to be discussed.  The first feature is that during each iteration the
algorithm should compute $a_0 = \sum_{k=1}^M a_k$ and then do the tau
leap only if the $\tau$ calculated via equation \eqref{eq:tau2} is
larger than some small multiple of $1/a_0$, but do one or more time
steps with an exact simulation method otherwise.  Switching between
tau-leaping and an exact method is reasonable because the benefits of
tau-leaping as compared with exact methods evaporate, and become
negative, as $\tau \to 1/a_0$, which is the expected amount of time
until the next reaction.

The second feature is more subtle.  For each leap, it is possible that
the leap condition will be violated so badly that some population
values will become negative. In fact, negative population values have
been found to occur in simulations using tau-leaping on systems of
interest.  \cite{Cao2005, Tian2004} As negative population values are
physically unreasonable, this constitutes a problem, and a number of
solutions have been proposed.  Tian et al.\cite{Tian2004} and
Chatterjee et al.  \cite{Chatterjee2005} independently developed a
method in which binomial random variables, as opposed to Poisson
random variables, are used to perform the leap.  Because binomial
random variables have bounded support, the parameters of the binomial
random variable can be chosen in a way that guarantees no molecular
species will become negative in the course of a leap.  Cao et al.
\cite{Cao2005} then developed a method to handle the potential of
negative population values in which the reactions are partitioned into
two sets before the calculation of $\tau$: critical reactions and
non-critical reactions.  For some predetermined integer, $n_c$,
between 2 and 20, the set of critical reactions is defined to consist
of those reactions with a positive propensity function that is within
$n_c$ firings of exhausting one of its reactants.  Having split the
reactions in such a way before a leap, the algorithm performs a
standard tau-leap for the non-critical reactions concurrent with a
standard Gillespie Algorithm step for the critical reactions.  It is
guaranteed that among all the critical reactions there will be at most
one firing during the leap, thereby significantly reducing, but not
completely doing away with, the chance of achieving a negative value.
If a negative population value is still achieved, the tau is shortened
and the leap is repeated.  See Ref. \onlinecite{Cao2005} for the full
details of this method.

While both the binomial tau-leaping method and the ``critical reaction
method'' guarantee negative population values will be avoided, neither
addresses the underlying problem of what is driving population values
negative: that the leap condition is badly violated at times.  Instead
of handling this larger problem, both the binomial tau-leap method and
the partitioning method of Cao et al. only handle it when species
numbers are low (although this is admittedly the most important time
to handle this problem).  Also, the fact that population values can
become negative in the absence of specific machinery designed to keep
them positive points out that other such large violations of the leap
condition are most likely occurring elsewhere in the simulation, yet
are going unnoticed.

\section{A new tau-leaping procedure}
\label{sec:newtau}

Through a post-leap check the procedure developed in this section will
only accept leaps that demonstrably satisfy a leap condition.  A
consequence of such enforcement will be that achieving negative
population values will be impossible, and so the partitioning
machinery of Cao et al.  will no longer be necessary.  Further, as the
method proposed will adaptively choose tau based upon the success or
failure of the previous leap, there will be no need to calculate tau
before each leap via equation \eqref{eq:tau2}.

\subsection{Conceptual framework}
\label{sec:concept}

The method proposed in this section relies heavily on the following
two facts: 1) the internal time frames of the Poisson processes are
distinct from each other and from the absolute time frame and 2) for
$T_2 \ge T_1 \ge T_k(t)$, the interval $Y_k(T_2) - Y_k(T_1)$ is
independent from the state of the system $X(t)$.  Consider equations
\eqref{eq:Rk} and \eqref{eq:state}.  Suppose that at time $t$ we have
knowledge of the state of the system, $X(t)$, the propensity
functions, $a_k = a_k(X(t))$, the various internal times $T_k = T_k(t)
= \int_0^t a_k(X(s))ds$, and the number of firings of each Poisson
process up to time $t$, $C_k \ \dot = \ Y_k(T_k(t))$.  However, we
suppose we have no information about the processes $Y_k(T) - Y_k(T_k)$
for $T > T_k$.  At this time we attempt to perform a leap with some
pre-determined $\tau$. By equation \eqref{eq:numfirings}, the number
of jumps of $Y_k$ over the internal time period $[T_k, T_k + a_k\tau)$
has a Poisson distribution with parameter $a_k\tau$.  We therefore
generate $M$ Poisson random variables and denote them by $N_k$.  Note
that we have now fixed the value $Y_k(T_k + a_k\tau) = N_k + C_k$ for
the course of the simulation.  We next approximate the state of the
system at time $t+\tau$ via equation \eqref{eq:update} and check the
leap condition.  If we verify that the leap condition has been
satisfied we may accept the updated system and attempt another leap.

If the leap condition is not satisfied we do not accept the leap, and
we do not update the system.  Instead, we decrease the tau value by
choosing some $\tau^* < \tau$ and attempt another leap over this
shorter time period.  However, we still know that $Y_k(T_k + a_k\tau)
= C_k + N_k$, for each $k$, and should condition upon this knowledge
when calculating $Y_k(T_k + a_k\tau^*)$.  Note that knowing $Y_k(T_k +
a_k\tau) = C_k + N_k$ is not the same as claiming to know that
reaction $k$ fired $C_k + N_k$ times by time $t + \tau$.  The former
equation is simply a statement about the values of the Poisson process
$Y_k(\cdot)$, while the latter is a statement about the actual firings
of the system by a certain time.  We prove the following theorem in
the appendix.

\begin{theorem}
  Let $Y(t)$ be a Poisson process with intensity $\lambda$, and let $0
  \le s < u < t$.  Then, conditioned on $Y(s)$ and $Y(t)$, $Y(u) -
  Y(s)$ has a $binomial(Y(t) - Y(s),r)$ distribution, where $r = (u -
  s)/(t-s)$.
  \label{thm:binomial}
\end{theorem}

\noindent By Theorem \ref{thm:binomial}, the distribution of $Y_k(T_k
+ a_k\tau^*) - Y_k(T_k)$, conditioned on $Y_k(T_k + a_k\tau) = C_k +
N_k$, has a binomial$(N_k, p_k)$ distribution, where $p_k =
\tau^*/\tau$.  After choosing the number of times $Y_k$ jumps in the
internal time period $[T_k, T_k + a_k\tau^*)$ according to the
binomial distribution just calculated we repeat the process of
attempting an update of the state of the system and of checking the
leap condition.  If this leap is also rejected, we simply store the
information of how many times $Y_k$ jumped by internal time $T_k +
a_k\tau^*$, shorten tau, and try again.  For the next attempted leap
we only need to condition on $Y_k(T_k)$ and $Y_k(T_k + a_k\tau^*)$
because of the independence of intervals of Poisson processes.
Eventually, a leap will be accepted and we may move forward in both
absolute and internal time.

We note that when we accept a leap and are ready to attempt another
one we may have stored the value of $Y_k(T)$ for many different
internal times, $T \ge T_k$.  When we attempt another leap, the next
proposed internal time will either fall between two internal times we
have stored, or will fall beyond our last stored internal time.  In
the former case, Theorem \ref{thm:binomial} may be applied because of
the independence of intervals of Poisson processes.  In the latter
case, the number of firings will be given as the number of firings up
to the last stored internal time, plus a Poisson random variable
accounting for the extra internal time.

The above description is the backbone of our new method.  At each
absolute time $t$ we will attempt a leap of size $\tau$.  Supposing
that we have stored the information $T^1,\dots, T^d$ and
$Y_k(T^1),\dots, Y_k(T^d)$, with $T^1,\dots,T^d > T_k$, $T_k +
a_k\tau$ will either fall between two of the stored internal times or
fall beyond $T^d$.  As above, in the case when $T^i \le T_k + a_k\tau
< T^{i+1}$ for some $i$, we apply Theorem \ref{thm:binomial} by
conditioning upon $Y_k(T^i)$ and $Y_k(T^{i+1})$ and choosing from the
appropriate binomial distribution.  Finally, we add the random
variable chosen to $Y_k(T^i) - C_k$ to find how many times $Y_k$
jumped between the internal times $T_k$ and $T_k + a_k\tau$.  In the
case when $T_k + a_k\tau \ge T^d$ we generate a Poisson random variable
with parameter $T_k + a_k\tau - T^d$ and add it to $Y_k(T^d) - C_k$ to
find the number of jumps.  Acceptance or rejection of the leap then
depends upon checking the leap condition.  If we do reject the leap,
we would then store the information just learned about each $Y_k$,
shorten $\tau$, and try again.  By storing the information gained
about the processes $Y_k$ after each attempted leap we see that no
information about the processes $Y_k$ will be lost in the course of a
simulation.  Because all of the randomness in the system resides in
the processes $Y_k$, we conclude that the rejection of leaps will not
skew the statistics of the sample paths.

We have not yet described how to update $\tau$ after each failed or
accepted leap and will do so now.  In the following we suppose that we
have already fixed the $\epsilon$ of the leap condition
\eqref{leapCond2} as $\overline \epsilon$.

\vspace{.1in}

\noindent \textbf{Tau updating procedure:}
\begin{enumerate}
\item If a leap is rejected because it fails the leap condition for
  $\epsilon = \overline \epsilon$, then decrease $\tau$ by multiplying
  it by some $p < 1$.
\item If a leap is accepted because it satisfied the leap condition
  for $\epsilon = \overline \epsilon$, but would have failed the leap
  condition if $ \epsilon = 3 \overline \epsilon/4$, then decrease
  $\tau$ by multiplying it by some $p^*$ that satisfies $p < p^* < 1$.
\item If a leap is accepted because it satisfied the leap condition
  for $\epsilon = \overline \epsilon$, and would have satisfied the
  leap condition if $ \epsilon = 3 \overline \epsilon/4$, then
  increase $\tau$ by raising it to the power $q$ for some $0< q < 1$.
\end{enumerate}

Unlike the method of Cao et al., we are not making an effort to select
the largest possible $\tau$ for which the leap condition will hold.
However, based upon the tau updating procedure given above, it should
be clear that we are attempting to select a tau that is at least near
such a maximal value.  However, as the value of $\epsilon$ itself is
rather arbitrary, it does not seem critical to select such a
``largest'' tau.

We point out that Step 2 in our tau updating procedure is useful to
keep the number of rejected leaps down.  In essence, Step 2 forces the
algorithm to always attempt to satisfy the leap condition for a
smaller value of epsilon than what was originally chosen, but does not
reject the leap if such a restrictive condition is not met.  Because
the generation of Poisson and binomial random variables are
computationally intensive procedures, attempting to limit the number
of failed leaps, and hence the number of random variables generated,
seems reasonable.

\subsection{The new algorithm}
\label{sec:newalg}

The analysis of the previous section gives us a new adaptive
tau-leaping procedure.  Before presenting the algorithm, however, some
notation is needed.  Each Poisson process $Y_k$ will have an
associated matrix, $S_k$, that will serve to store the information
gained from leaps that fail the post leap check. Each $S_k$ has two
columns. The first column will store internal times (as opposed to
absolute times).  The second column will store the number of firings
of $Y_k$ up to the internal time in the first column.  That is, the
elements of row $i$ satisfy $Y_k(S_k(i,1)) = S_k(i,2)$.  The first row
of $S_k$ will always contain the present internal time and the number
of times $Y_k$ has fired up to that time.  Also, $T_k = T_k(t)$ will
always denote the current internal time of $Y_k$ and $Y_k(T_k)$ will
be denoted by $C_k$. Combining the above gives that at each step we
have $T_k = S_k(1,1)$ and $C_k = S_k(1,2)$.  Finally, the values
$row_k$ will be used to update the rows of the matrices $S_k$ after
every step.

\begin{algorithm} (Post-leap check tau-leaping)
  \begin{enumerate}
    \setlength{\itemsep}{1pt} \setlength{\parskip}{0pt}
    \setlength{\parsep}{0pt}
  \item Initialize.  Set the initial number of molecules of each
    species, $x \in \Z^N_{\ge 0}$, and calculate the propensity
    functions, $a_k$.  Set $t = 0$ and for each $k$ set $T_k = C_k =
    0$, and $S_k = [0 \ , \ 0]$.  Calculate $\tau$ via equation
    \eqref{eq:tau2}.  Set $0 < p < p^* < 1$ and $0< q < 1$.
  \item For each $k$ do the following:
    \label{return1}
    \begin{enumerate}
    \item Let $B_k = $ the number of rows of $S_k$.
    \item If $a_k \tau + T_k \ge S_k(B_k,1)$,
      \begin{itemize}
      \item Set $N_k = \hbox{Poisson}(T_k + a_k\tau - S_k(B_k,1)) +
        S_k(B_k,2) - C_k$.
      \item Set $row_k = B_k$.
      \end{itemize}
    \item else
      \begin{itemize}
      \item Find the index, $I_k$, such that
        \begin{align*}
          S_k(I_k - 1,1) \le T_k + a_k \tau < S_k(I_k,1).
        \end{align*}
      \item Set $r = (T_k + a_k\tau - S_k(I_k - 1,1)) / \left(
          S_k(I_k, 1) - S_k(I_k-1,1) \right).$
      \item Set $N_k = \hbox{binomial}(S_k(I_k,2) - S_k(I_k - 1,2),r)
        + S_k(I_k - 1,2) - C_k$.
      \item Set $row_k = I_k - 1$.
      \end{itemize}
    \end{enumerate}
  \item Check whether the leap condition holds with the selected
    $N_k$.
  \item If yes, accept leap:
    \begin{enumerate}
    \item Update each $S_k$.
      \label{succ:update}
      \begin{itemize}
      \item Delete all rows less than or equal to $row_k$ and shift
        all other rows down.  Add a new first row of $[T_k + a_k\tau \
        , \ C_k + N_k]$.
      \end{itemize}
    \item Set $t = t + \tau$.
    \item For each $k$, set $T_k = T_k + a_k\tau$ and $C_k = C_k +
      N_k$.
    \item Update $\tau$ according to the tau updating procedure of the
      previous section.
    \item Set $x = x + \sum_{k=1}^M N_k(\nu_k' - \nu_k)$ and
      recalculate the propensity functions.
    \item Return to step \ref{return1}.
    \end{enumerate}

  \item Else, reject leap:
    \begin{enumerate}
    \item Update each $S_k$.
      \label{rej:update}
      \begin{itemize}
      \item Add the row $[T_k + a_k\tau \ , \ C_k + N_k]$ between rows
        $row_k$ and $row_k + 1$ (if $row_k + 1 > B_k$, just add a last
        row to $S_k$).
      \end{itemize}
    \item Decrease $\tau$ by setting $\tau = p\tau$.
      \label{lowertau2}
    \item Return to step \ref{return1}.
    \end{enumerate}
  \end{enumerate}
  \label{alg:fulltau}
\end{algorithm}

\begin{remark}
  In the above algorithm no population value can become negative after
  an accepted leap.  Unlike the binomial tau-leap method or the
  splitting of the reactions into critical and non-critical subsets,
  however, Algorithm \ref{alg:fulltau} handles the underlying problem
  that could cause negative population values.  That is, the leap
  condition will never be violated.
\end{remark}

We note that the manner in which we choose our $\tau$ will generally
lead to smaller $\tau$ values than the pre-leap computation method for
a given value of $\epsilon$ and for a given state of the system
$X(t)$.  Therefore, for a given $\epsilon$ we expect that our method
will need more simulation time than Algorithm \ref{alg:oldtau}, but
will produce more accurate results.  Thus, in order to find which
algorithm is more efficient we will need to compare them with
different $\epsilon$ values.  Also, we note that Algorithm
\ref{alg:oldtau} chooses its tau values based upon the current state
of the system whereas Algorithm \ref{alg:fulltau} chooses its tau
values based upon the success or failure of the previously attempted
leap.  While it is true that each method will (at least statistically)
produce the same leap for a given state of the system and a given
$\tau$, we note that over the course of an entire simulation the
difference in how each algorithm selects their tau values will cause
the statistics of the sample paths to diverge.  Therefore it is
entirely plausible that one method will achieve higher accuracy than
the other through fewer steps.  This is demonstrated in Section
\ref{sec:dimer}.

\subsection{Switching to an exact algorithm}
\label{sec:switch}

As noted in the paragraph following Algorithm \ref{alg:oldtau}, it is
sometimes necessary to switch between a tau-leaping method and an
exact method.  Doing so for Algorithm \ref{alg:fulltau} is non-trivial
as there could be stored future information in the matrices $S_k$.  If
any of the $S_k$ matrices do have stored future information, then the
distributions of the future states of the system do not solely depend
upon the current state of the system.  A choice must therefore be made
as to how to make the switch in Algorithm \ref{alg:fulltau}.  One
option is to discard all stored future information (that is, delete
the information in the matrices $S_k$) and switch to an exact method.
In this case, it is important to recognize that by discarding the
stored future information, we have, in effect, changed the Poisson
processes, $Y_k$, and such a change adds a potential bias to the
choice of sample paths.  For example, it is possible that the
algorithm needed to switch to an exact method because one or more of
the processes $Y_k$ had significantly more firings than was expected
over a short period of internal time.  By discarding this stored
future information, we may be inadvertently biasing the system away
from such instances.  Typically, however, there are not many switches
made from tau-leaping to an exact method in the course of a
simulation, and so the bias that is added may be negligible.  Further,
doing so would be very simple to implement.

The other option is to keep all of the stored future information and
still switch to an exact method.  In this case, we preserve the fact
that we are not approximating the processes $Y_k$ in our simulation.
The exact method to which we will switch is similar to the modified
Next Reaction Method. See Ref.  \onlinecite{Anderson2007a} for a
detailed explanation of the modified Next Reaction Method.

As in the modified Next Reaction Method, we begin by letting $P_k$
denote the internal time of the next firing of the Poisson process
$Y_k$.  That is, for each $k$ we let $P_k = \min\{T > T_k \ | \ Y_k(T)
> Y_k(T_k)\}$.  There are two cases to consider in the calculation of
$P_k$:

\noindent \textbf{Case 1:} If there is no stored information for the
$k$th reaction we may use the fact that the $Y_k$'s are independent,
unit rate Poisson processes and set $P_k = T_k + \ln(1/r_k)$, where
$r_k$ is uniform$(0,1)$.

\noindent \textbf{Case 2:} If there is stored information for the
$k$th reaction channel we must condition upon that information in
order to calculate the distribution of the next firing time.  In the
Appendix we show the following.
\begin{lemma}
  Let $Y(t)$ be a Poisson process with intensity $\lambda$.  Let $t >
  0$ and $0 = Y(0) < Y(t) = N$.  Let $P^1 = \min\{ s \ | \ Y(s) >
  0\}$.  Then, $P(P^1 > r \ | \ Y(t) = N) = (1 - r/t)^N$.
  \label{lemma:findP}
\end{lemma}

We let $B_k$, $C_k$, and $T_k$ be as in Algorithm \ref{alg:fulltau}.
By Lemma \ref{lemma:findP} we may calculate $P_k$ in the following
manner.  First, find the index $j$ such that $C_k = S_k(j,2)$ and $C_k
< S_k(j+1,2)$.  Thus, the next firing of $Y_k$ happens in the internal
time interval $[S_k(j,1),S_k(j+1,1))$.  If no such $j$ exists, set
$P_k = S_k(B_k,1) + \ln(1/r_k)$, where $r_k$ is uniform$(0,1)$.  If
such a $j$ does exist, let $t_k = S_k(j+1,1) - S_k(j,1)$ and let $N_k
= S_k(j+1,2) - S_k(j,2)$.  $N_k$ gives the number of firings of the
Poisson process $Y_k$ in the internal time period
$[S_k(j,1),S_k(j+1,1))$, which is of length $t_k$.  By Lemma
\ref{lemma:findP}, $P_k - S_k(j,1)$ has distribution function $(1 -
r/t_k)^{N_k}$ and so we may set $P_k = S_k(j,1) + t_k( 1 -
r_k^{1/N_k})$, where $r_k$ is uniform$(0,1)$.

%\singlespacing
\begin{algorithm} An exact stochastic simulation algorithm given stored
  future information.
  \begin{enumerate}
    \setlength{\itemsep}{1pt}
    \setlength{\parskip}{0pt}
    \setlength{\parsep}{0pt}
  \item Input: the number of molecules of each species, $t$, and for
    each $k$, the following from Algorithm \ref{alg:fulltau}: $S_k$,
    $T_k$, and $C_k$.
  \item For each $k$, find $P_k$ as described in the previous
    paragraph.
  \item Set $\Delta_k = (P_k - T_k)/a_k$.
    \label{SSAstep2}
  \item Set $\Delta = \min\{\Delta_k\}$ and let $\Delta_{\mu}$ be the
    value at which the minimum is realized.
  \item Set $t = t + \Delta$ and update the number of each
    molecular species according to reaction $\mu$.
  \item For each $k$, set $T_k = T_k + a_k\Delta$.
  \item Set $C_{\mu} = C_{\mu} + 1$.
  \item Find $P_{\mu}$ as described in the previous paragraph.
  \item Recalculate the propensity functions.
  \item Update each $S_k$ by deleting all rows with internal times
    less than or equal to $T_k$ and adding a new first row of $[T_k \
    , \ C_k]$.
  \item Return to step \ref{SSAstep2} or return to tau-leaping.
  \end{enumerate}
  \label{alg:fulltauSSA}
\end{algorithm}

%\doublespacing
\begin{remark}
  After the first time step, the above algorithm uses only one random
  variable per time step.  Also, Algorithm \ref{alg:fulltauSSA}
  becomes the modified Next Reaction Method if all of the information
  in each $S_k$ is exhausted before the switch back to tau-leaping is
  made.
\end{remark}

We have shown how to switch successfully from tau-leaping with
Algorithm \ref{alg:fulltau} to the exact method in Algorithm
\ref{alg:fulltauSSA}.  However, we now have to consider how to switch
back.  It is not instantly clear that we can simply discard the
information contained in the unused $P_k$ values without adding bias
to our system.  We also note that we can not simply incorporate the
$P_k$ values into the $S_k$ matrices because each $S_k$ contains
information about how many jumps of $Y_k$ have taken place up to
different internal times and not information about the {\em exact}
jump times.  However, the following theorem allows us to discard the
information stored in the $P_k$ values when we switch from Algorithm
\ref{alg:fulltauSSA} back to Algorithm \ref{alg:fulltau}.  The proof
can be found in the appendix.

\begin{theorem}
  The statistics of the firing times of each reaction channel are
  unaffected by discarding the $P_k$ values when we switch from the
  exact method of Algorithm \ref{alg:fulltauSSA} to tau-leaping in
  Algorithm \ref{alg:fulltau}.
  \label{thm:switch}
\end{theorem}

\section{A numerical example}
\label{sec:dimer}

We compare the different tau-leaping methods on a model of an unstable
dimer that has been used in a number of earlier papers.
\cite{Gill2001, Gill2003, Cao2006} The model consists of four
reactions and three species.  The reactions are
\begin{equation}
  \begin{array}{l l}
    X_1 \xrightarrow{c_1} 0 \ \ \ &  X_2 \xrightarrow{c_3} 2X_1 \\
    2X_1 \xrightarrow{c_2} X_2 \ \ \ &  X_2 \xrightarrow{c_4} X_3,
  \end{array}
  \label{dimer}
\end{equation}
with rate constant $c_1 = 1, \ c_2 = .002, \ c_3 = 0.5,$ and $c_4=
0.04$.  The propensity functions are $a_1(X) = X_1, \ a_2(X) =
(.002/2)X_1(X_1 - 1), \ a_3(X) = 0.5X_2,$ and $a_4(X) = 0.04X_2$.  We
chose initial conditions of $X_1(0) = 10^6$, $X_2(0) = 10^3$, and
$X_3(0) = 0$.  We imposed the leap condition \eqref{leapCond2} in
Algorithm \ref{alg:oldtau} by calculating tau before each leap
according to equation \eqref{eq:tau2} and imposed the leap condition
in Algorithm \ref{alg:fulltau} by checking that the condition was
satisfied after every leap.  For Algorithm \ref{alg:fulltau} we chose
$p = .75$, $p^* = .9$, and $q = .98$ as the values for our tau
updating procedure described at the end of Section \ref{sec:concept}.

For $\epsilon = 0.17$, $\epsilon = 0.1$, and $\epsilon = 0.03$ as the
$\epsilon$ values of the leap condition \eqref{leapCond2}, we
simulated the above system $10^4$ times using Gillespie's original
algorithm (SSA), Algorithm \ref{alg:oldtau}, and Algorithm
\ref{alg:fulltau}.  The simulations were performed using Matlab on a 2
Ghz processor running on the Debian operating system.  To generate
Poisson and binomial random variables, we used the Matlab Poisson and
binomial random number generators.  Each simulation began at time
$t=0$ and ended at time $t=1$.  In Figure 1
                                                          %%\ref{fig:lots}
we show the histograms of $X_1(1)$ and $X_2(1)$ for the different
values of $\epsilon$.  We see that for each $\epsilon$ Algorithm
\ref{alg:fulltau} is significantly more accurate than Algorithm
\ref{alg:oldtau}.  As was pointed out at the end of Section
\ref{sec:newalg}, however, the tau selection strategy for Algorithm
\ref{alg:fulltau} will naturally choose smaller values of $\tau$ for a
given value of $\epsilon$.  Therefore, it is not surprising that
Algorithm \ref{alg:fulltau} is significantly more accurate for each
$\epsilon$.  The CPU times for the different methods and different
$\epsilon$ values are given in the following table.

\vspace{.125in}
\begin{center}
  \begin{tabular}{|l| c|c|c|}
    \hline
    CPU Times \ & $\epsilon = 0.17$ & $\epsilon = 0.10$ & $\epsilon =
    0.03$ \\
    \hline
    Alg. \ref{alg:oldtau} & \ 31 CPU Minutes \ & \  46 CPU Minutes \
    & \  120 CPU Minutes \  \\
    Alg. \ref{alg:fulltau} & \ 60 CPU Minutes \ & \  82 CPU Minutes \
    & \  226 CPU Minutes  \ \\
    \hline
  \end{tabular}
\end{center}
\vspace{.125in}

\noindent As was predicted in Section \ref{sec:newalg}, the increased
accuracy of Algorithm \ref{alg:fulltau} for each $\epsilon$ came at
the price of longer CPU times.

In Figure 2 %\ref{fig:compare}
we plot the histograms of $X_1(1)$ and $X_2(1)$ of the Gillespie
Algorithm (SSA), Algorithm \ref{alg:oldtau} with $\epsilon = 0.03$,
and Algorithm \ref{alg:fulltau} with $\epsilon = 0.1$.  We see that
the histograms of the different tau-leaping methods with their
respective $\epsilon$ values are now nearly equivalent in their
accuracy.  However, based upon the CPU times in the above table, in
order to get such accuracy Algorithm \ref{alg:oldtau} required 46\%
more CPU time than Algorithm \ref{alg:fulltau}.  We therefore see
that, at least for this model, our new post-leap check method is more
efficient than the pre-leap computation method.

We next simulated the system 100 times using Algorithm
\ref{alg:oldtau} with $\epsilon = 0.03$ and Algorithm
\ref{alg:fulltau} with $\epsilon = 0.1$, except this time we added
machinery to keep track of how many leaps were needed in each
algorithm to complete the simulation.  We found that Algorithm
\ref{alg:fulltau} took an average of 203.4 successful leaps per
simulation and rejected an average of 37.8 leaps.  We found that
Algorithm \ref{alg:oldtau} took an average of 466.8 leaps.  Therefore,
Algorithm \ref{alg:fulltau} needed 56\% fewer successful leaps than
Algorithm \ref{alg:oldtau}, which implies that the average size of tau
for the successful leaps of Algorithm \ref{alg:fulltau} was
approximately double that of Algorithm \ref{alg:oldtau}.  The reason
Algorithm \ref{alg:fulltau} can achieve similar accuracy to Algorithm
\ref{alg:oldtau} with larger average tau values was explained at the
end of Section \ref{sec:newalg}.

\section{Discussion}

By explicitly representing the reaction times of discrete stochastic
chemical systems as the firing times of independent, unit rate Poisson
processes, we have developed an accurate and efficient adaptive
tau-leaping procedure.  The main difference between the method
developed in this paper and the current adaptive tau-leaping methods
is that we enforce our leap conditions via a post-leap check.
%, as opposed to computing a tau that will result in a leap that will satisfy the leap condition with some probability via a procedure before each leap.  
Also, we have demonstrated how to reject leaps without affecting the
statistics of the sample paths generated.  Further, as a consequence
of always satisfying a given leap condition, our procedure is
guaranteed to never produce negative population values, which is in
contrast to current methods in which extra machinery is needed to
guarantee that population values remain non-negative.  Finally,
through an example of an unstable dimer, we have demonstrated that the
method proposed in this paper is not only extremely accurate, but is
also extremely efficient.

Algorithm \ref{alg:fulltau} will surely not be more efficient than
Algorithm \ref{alg:oldtau} for all chemical reaction systems.
However, by enforcing a post-leap check in such a way that the
statistics of the sample paths are not skewed by the rejection of a
leap, we are confident that Algorithm \ref{alg:fulltau} will be more
accurate and will have better stability properties for all chemical
reaction systems.  We also note that the method we have developed is
easily adaptable to any leap conditions, not just those given by
\eqref{leapCond1} and \eqref{leapCond2}.  Such adaptability is in
contrast to methods that compute $\tau$ before each leap, which need a
new tau computation procedure for every new leap condition.

This paper represents only a first step in developing and analyzing
tau-leaping methods through an understanding of the random time change
representation given in equation \eqref{eq:state}.  Future work will
focus on error analysis and the development of methods that will
achieve greater accuracy through fewer steps.

\vspace{.25in}
\begin{center}
  {\bf \large Acknowledgments}
\end{center}

I would like to thank Thomas G. Kurtz for enlightening conversations
that helped in the development of this work.  I would also like to
thank an anonymous reviewer for detailed, helpful comments that
improved the clarity of this paper.  This work was done under the
support of NSF grant DMS-0553687.

\appendix

\section{Definition of functions}
\label{app:g}

The functions $g_i = g_i(X(t))$ used in equation \eqref{eq:tau2} are
defined as follows.  Let HOR$(i)$ denote the highest order of reaction
in which species $X_i$ appears as a reactant.
\begin{enumerate}[(i)]
  \setlength{\itemsep}{1pt}
  \setlength{\parskip}{0pt}
  \setlength{\parsep}{0pt}
\item If HOR$(i) = 1$, take $g_i = 1$.
\item If HOR$(i) = 2$, take $g_i=2$, except if any second-order
  reaction requires two $X_i$ molecules in which case take $g_i = 2 +
  1/(X_i(t) - 1)$.
\item If HOR$(i) = 3$, take $g_i = 3$, except if some third-order
  reaction requires two $X_i$ molecules in which case take
  \begin{equation*}
    g_i = \frac{3}{2}\left(2 + \frac{1}{X_i(t)-1}\right),
  \end{equation*}
  except if some third-order reaction requires three $X_i$ molecules
  in which case take
  \begin{equation*}
    g_i = 3 + \frac{1}{X_i(t)-1} + \frac{2}{X_i(t) - 2}.
  \end{equation*}
\end{enumerate}

\section{Proofs}
\label{app:proofs}

\begin{proof}(of Theorem \ref{thm:binomial}) Without loss of
  generality, we suppose that $s = 0$ and $Y(0) = 0$.  Let $Y(t) =
  N$ and $0 < u < t$. Then
  \begin{align*}
    P(Y(u) = j \ | \ Y(t) = N) & = P(Y(u) = j, Y(t) = N)/P(Y(t) = N)\\
    &= P(Y(t) - Y(u) = N - j)P(Y(u) = j)/P(Y(t) = N)\\
    &= \frac{e^{-\lambda(t-u)}(\lambda(t-u))^{N-j}}
    {(N-j)!}\frac{e^{-\lambda u} (\lambda u)^j}{j!} \frac{N!}
    {e^{-\lambda t}(\lambda t)^N}\\
    &= \binom{N}{j}\left(\frac{u}{t}\right)^j\left(1 -
      \frac{u}{t}\right)^{N-j}.
  \end{align*}
\end{proof}

\begin{proof}(of Lemma \ref{lemma:findP})
  \begin{align*}
    P(P^1 > r \ | \ Y(t) = N) &= P(Y(r) = 0 \ | \ Y(t) = N)\\
    &= P(Y(r) = 0, Y(t) = N)/P(Y(t) = N)\\
    &= P(Y(t) - Y(r) = N)P(Y(r) = 0)/P(Y(t) = N)\\
    &= \frac{e^{-\lambda(t-r)}(\lambda(t - r))^N}{N!}e^{-\lambda
      r}\frac{N!}{e^{-\lambda t}(\lambda t)^N}\\
    &= (1 - r/t)^N.
  \end{align*}
\end{proof}

\begin{proof}(of Theorem \ref{thm:switch}) If there are no stored
  reaction times for reaction channel $k$, then discarding the extra
  information contained in $P_k$ is done by invoking the loss of
  memory property.  Now suppose that there is stored information for
  reaction $k$.  The firing times less than $T>0$ of a Poisson process
  $Y$, conditioned on $Y(T)$, are uniform$(0,T)$ random variables.
  Therefore, it is sufficient to show that for a uniform$(a,b)$ random
  number $U$, $P(U > s + x \ | \ U > x) = (b-(s+x))/(b-x)$, thus
  showing that the conditional statistics are uniform$(x,b)$.  The
  calculation is simple and is omitted.
\end{proof}

\bibliography{AndTauRev2}

\begin{thebibliography}{18}
\expandafter\ifx\csname natexlab\endcsname\relax\def\natexlab#1{#1}\fi
\expandafter\ifx\csname bibnamefont\endcsname\relax
  \def\bibnamefont#1{#1}\fi
\expandafter\ifx\csname bibfnamefont\endcsname\relax
  \def\bibfnamefont#1{#1}\fi
\expandafter\ifx\csname citenamefont\endcsname\relax
  \def\citenamefont#1{#1}\fi
\expandafter\ifx\csname url\endcsname\relax
  \def\url#1{\texttt{#1}}\fi
\expandafter\ifx\csname urlprefix\endcsname\relax\def\urlprefix{URL }\fi
\providecommand{\bibinfo}[2]{#2}
\providecommand{\eprint}[2][]{\url{#2}}

\bibitem[{\citenamefont{Gillespie}(1976)}]{Gill76}
\bibinfo{author}{\bibfnamefont{D.~T.} \bibnamefont{Gillespie}},
  \bibinfo{journal}{J. Comput. Phys.} \textbf{\bibinfo{volume}{22}},
  \bibinfo{pages}{403} (\bibinfo{year}{1976}).

\bibitem[{\citenamefont{Gillespie}(1977)}]{Gill77}
\bibinfo{author}{\bibfnamefont{D.~T.} \bibnamefont{Gillespie}},
  \bibinfo{journal}{J. Phys. Chem.} \textbf{\bibinfo{volume}{81}},
  \bibinfo{pages}{2340} (\bibinfo{year}{1977}).

\bibitem[{\citenamefont{Andersson and Britton}(2000)}]{Andersson2000}
\bibinfo{author}{\bibfnamefont{H.}~\bibnamefont{Andersson}} \bibnamefont{and}
  \bibinfo{author}{\bibfnamefont{T.}~\bibnamefont{Britton}},
  \emph{\bibinfo{title}{Stochastic Epidemic Models and Their Statistical
  Analysis}} (\bibinfo{publisher}{Springer Lectures Notes in Statistics},
  \bibinfo{address}{Berlin}, \bibinfo{year}{2000}).

\bibitem[{\citenamefont{Anderson}(2007)}]{Anderson2007a}
\bibinfo{author}{\bibfnamefont{D.~F.} \bibnamefont{Anderson}},
  \bibinfo{journal}{J. Chem. Phys.} \textbf{\bibinfo{volume}{127}},
  \bibinfo{pages}{214107} (\bibinfo{year}{2007}).

\bibitem[{\citenamefont{Ball et~al.}(2006)\citenamefont{Ball, Kurtz, Popovic,
  and Rempala}}]{Ball06}
\bibinfo{author}{\bibfnamefont{K.}~\bibnamefont{Ball}},
  \bibinfo{author}{\bibfnamefont{T.~G.} \bibnamefont{Kurtz}},
  \bibinfo{author}{\bibfnamefont{L.}~\bibnamefont{Popovic}}, \bibnamefont{and}
  \bibinfo{author}{\bibfnamefont{G.}~\bibnamefont{Rempala}},
  \bibinfo{journal}{Ann. Appl. Prob.} \textbf{\bibinfo{volume}{16}},
  \bibinfo{pages}{1925} (\bibinfo{year}{2006}).

\bibitem[{\citenamefont{Kurtz}(1977/78)}]{Kurtz78}
\bibinfo{author}{\bibfnamefont{T.~G.} \bibnamefont{Kurtz}},
  \bibinfo{journal}{Stoch. Proc. Appl.} \textbf{\bibinfo{volume}{6}},
  \bibinfo{pages}{223} (\bibinfo{year}{1977/78}).

\bibitem[{\citenamefont{Ethier and Kurtz}(1986)}]{Kurtz86}
\bibinfo{author}{\bibfnamefont{S.~N.} \bibnamefont{Ethier}} \bibnamefont{and}
  \bibinfo{author}{\bibfnamefont{T.~G.} \bibnamefont{Kurtz}},
  \emph{\bibinfo{title}{Markov Processes: Characterization and Convergence}}
  (\bibinfo{publisher}{John Wiley \& Sons}, \bibinfo{address}{New York},
  \bibinfo{year}{1986}).

\bibitem[{\citenamefont{Gibson and Bruck}(2000)}]{Gibson2000}
\bibinfo{author}{\bibfnamefont{M.}~\bibnamefont{Gibson}} \bibnamefont{and}
  \bibinfo{author}{\bibfnamefont{J.}~\bibnamefont{Bruck}}, \bibinfo{journal}{J.
  Phys. Chem. A} \textbf{\bibinfo{volume}{105}}, \bibinfo{pages}{1876}
  (\bibinfo{year}{2000}).

\bibitem[{\citenamefont{Arkin et~al.}(1998)\citenamefont{Arkin, Ross, and
  McAdams}}]{Arkin1998}
\bibinfo{author}{\bibfnamefont{A.}~\bibnamefont{Arkin}},
  \bibinfo{author}{\bibfnamefont{J.}~\bibnamefont{Ross}}, \bibnamefont{and}
  \bibinfo{author}{\bibfnamefont{H.~H.} \bibnamefont{McAdams}},
  \bibinfo{journal}{Genetics} \textbf{\bibinfo{volume}{149}},
  \bibinfo{pages}{1633} (\bibinfo{year}{1998}).

\bibitem[{\citenamefont{McAdams and Arkin}(1997)}]{McAdams1997}
\bibinfo{author}{\bibfnamefont{H.~H.} \bibnamefont{McAdams}} \bibnamefont{and}
  \bibinfo{author}{\bibfnamefont{A.}~\bibnamefont{Arkin}},
  \bibinfo{journal}{Proc. Natl. Acad. Sci.} \textbf{\bibinfo{volume}{94}},
  \bibinfo{pages}{814} (\bibinfo{year}{1997}).

\bibitem[{\citenamefont{Ozbudak et~al.}(2002)\citenamefont{Ozbudak, Thattai,
  Kurtser, Grossman, and van Oudenaarden}}]{Ozbudak2002}
\bibinfo{author}{\bibfnamefont{E.~M.} \bibnamefont{Ozbudak}},
  \bibinfo{author}{\bibfnamefont{M.}~\bibnamefont{Thattai}},
  \bibinfo{author}{\bibfnamefont{I.}~\bibnamefont{Kurtser}},
  \bibinfo{author}{\bibfnamefont{A.~D.} \bibnamefont{Grossman}},
  \bibnamefont{and} \bibinfo{author}{\bibfnamefont{A.}~\bibnamefont{van
  Oudenaarden}}, \bibinfo{journal}{Nat. Genet.} \textbf{\bibinfo{volume}{31}},
  \bibinfo{pages}{69} (\bibinfo{year}{2002}).

\bibitem[{\citenamefont{Samad et~al.}(2005)\citenamefont{Samad, Khammash,
  Petzold, and Gillespie}}]{Petzold2005}
\bibinfo{author}{\bibfnamefont{H.~E.} \bibnamefont{Samad}},
  \bibinfo{author}{\bibfnamefont{M.}~\bibnamefont{Khammash}},
  \bibinfo{author}{\bibfnamefont{L.}~\bibnamefont{Petzold}}, \bibnamefont{and}
  \bibinfo{author}{\bibfnamefont{D.}~\bibnamefont{Gillespie}},
  \bibinfo{journal}{Inter. J. Robust Nonlin. Contr.}
  \textbf{\bibinfo{volume}{15}}, \bibinfo{pages}{691} (\bibinfo{year}{2005}).

\bibitem[{\citenamefont{Gillespie}(2001)}]{Gill2001}
\bibinfo{author}{\bibfnamefont{D.~T.} \bibnamefont{Gillespie}},
  \bibinfo{journal}{J. Chem. Phys.} \textbf{\bibinfo{volume}{115}},
  \bibinfo{pages}{1716} (\bibinfo{year}{2001}).

\bibitem[{\citenamefont{Cao et~al.}(2006)\citenamefont{Cao, Gillespie, and
  Petzold}}]{Cao2006}
\bibinfo{author}{\bibfnamefont{Y.}~\bibnamefont{Cao}},
  \bibinfo{author}{\bibfnamefont{D.~T.} \bibnamefont{Gillespie}},
  \bibnamefont{and} \bibinfo{author}{\bibfnamefont{L.~R.}
  \bibnamefont{Petzold}}, \bibinfo{journal}{J. Chem. Phys.}
  \textbf{\bibinfo{volume}{124}}, \bibinfo{pages}{044109}
  (\bibinfo{year}{2006}).

\bibitem[{\citenamefont{Cao et~al.}(2005)\citenamefont{Cao, Gillespie, and
  Petzold}}]{Cao2005}
\bibinfo{author}{\bibfnamefont{Y.}~\bibnamefont{Cao}},
  \bibinfo{author}{\bibfnamefont{D.~T.} \bibnamefont{Gillespie}},
  \bibnamefont{and} \bibinfo{author}{\bibfnamefont{L.~R.}
  \bibnamefont{Petzold}}, \bibinfo{journal}{J. Chem. Phys.}
  \textbf{\bibinfo{volume}{123}}, \bibinfo{pages}{054104}
  (\bibinfo{year}{2005}).

\bibitem[{\citenamefont{Tian and Burrage}(2004)}]{Tian2004}
\bibinfo{author}{\bibfnamefont{T.}~\bibnamefont{Tian}} \bibnamefont{and}
  \bibinfo{author}{\bibfnamefont{K.}~\bibnamefont{Burrage}},
  \bibinfo{journal}{J. Chem. Phys.} \textbf{\bibinfo{volume}{121}},
  \bibinfo{pages}{10356} (\bibinfo{year}{2004}).

\bibitem[{\citenamefont{Chatterjee and Vlachos}(2005)}]{Chatterjee2005}
\bibinfo{author}{\bibfnamefont{A.}~\bibnamefont{Chatterjee}} \bibnamefont{and}
  \bibinfo{author}{\bibfnamefont{D.~G.} \bibnamefont{Vlachos}},
  \bibinfo{journal}{J. Chem. Phys.} \textbf{\bibinfo{volume}{122}},
  \bibinfo{pages}{024112} (\bibinfo{year}{2005}).

\bibitem[{\citenamefont{Gillespie and Petzold}(2003)}]{Gill2003}
\bibinfo{author}{\bibfnamefont{D.~T.} \bibnamefont{Gillespie}}
  \bibnamefont{and} \bibinfo{author}{\bibfnamefont{L.~R.}
  \bibnamefont{Petzold}}, \bibinfo{journal}{J. Chem. Phys.}
  \textbf{\bibinfo{volume}{119}}, \bibinfo{pages}{8229} (\bibinfo{year}{2003}).

\end{thebibliography}

\pagebreak
\noindent Figures

Figure 1%\ref{fig:lots}

\vspace{1in}

 \begin{center}
   \includegraphics[width = 2.9in,height=2.2in]{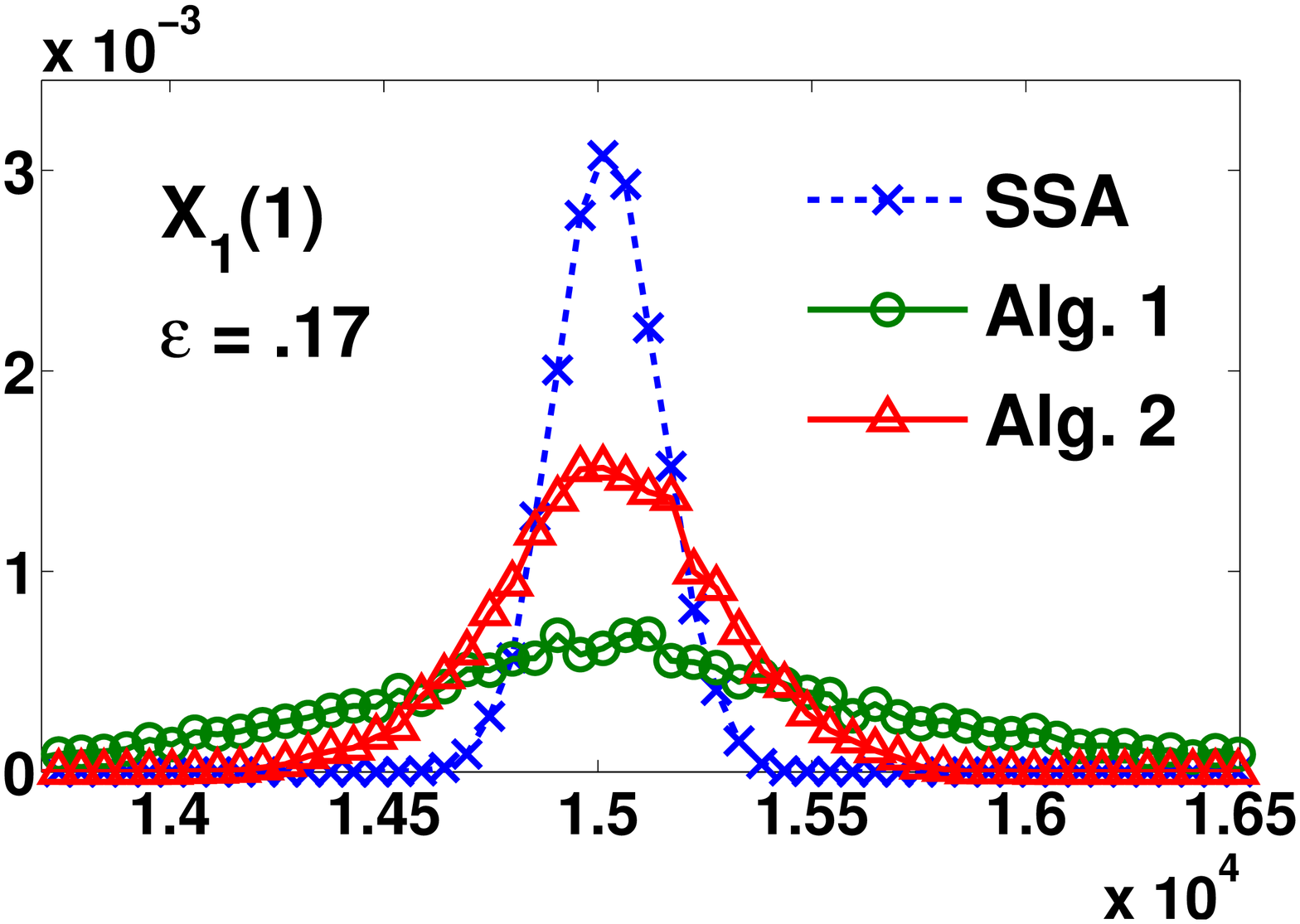}
   \includegraphics[width = 2.9in,height=2.2in]{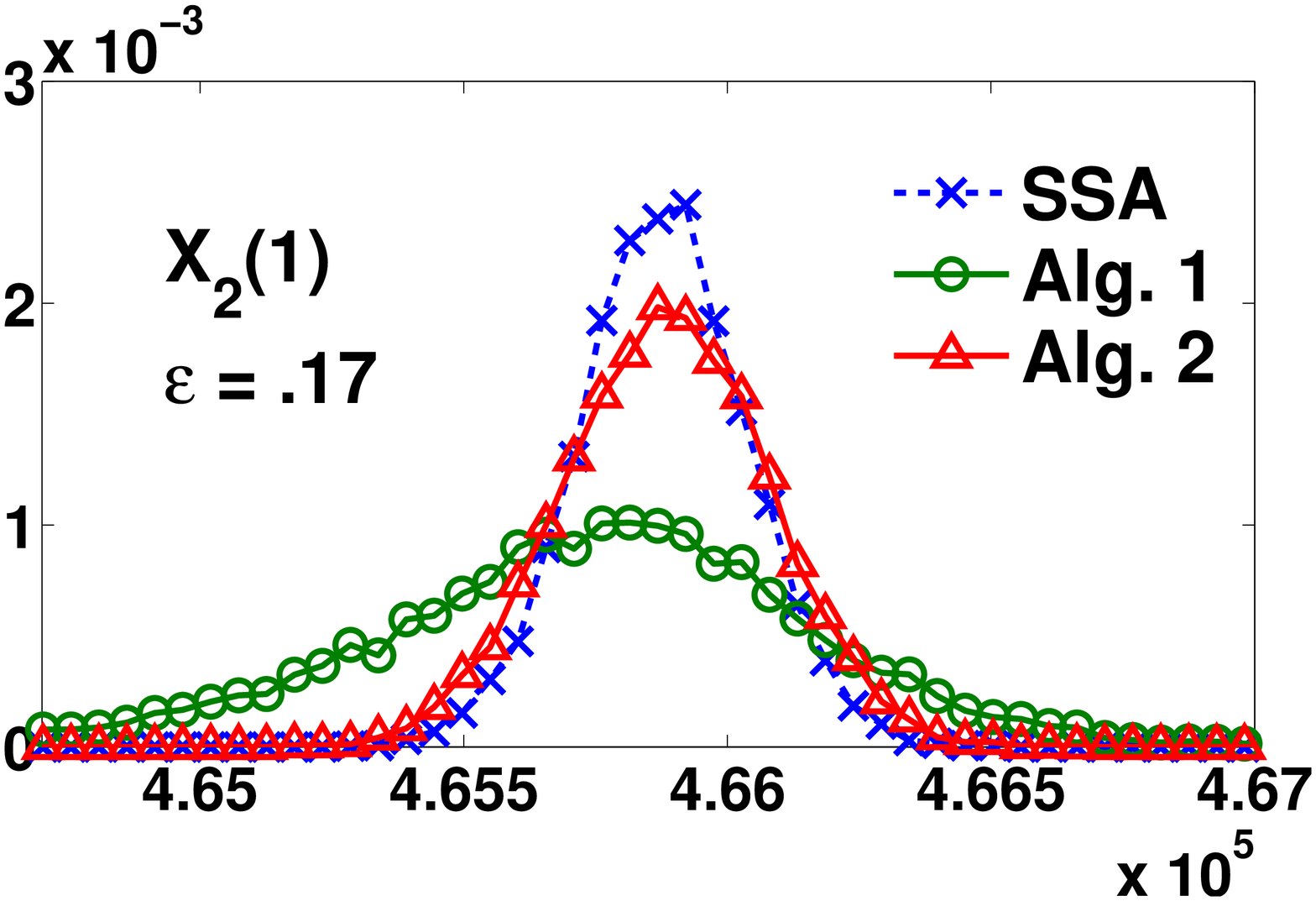}\\
    \includegraphics[width = 2.9in,height=2.2in]{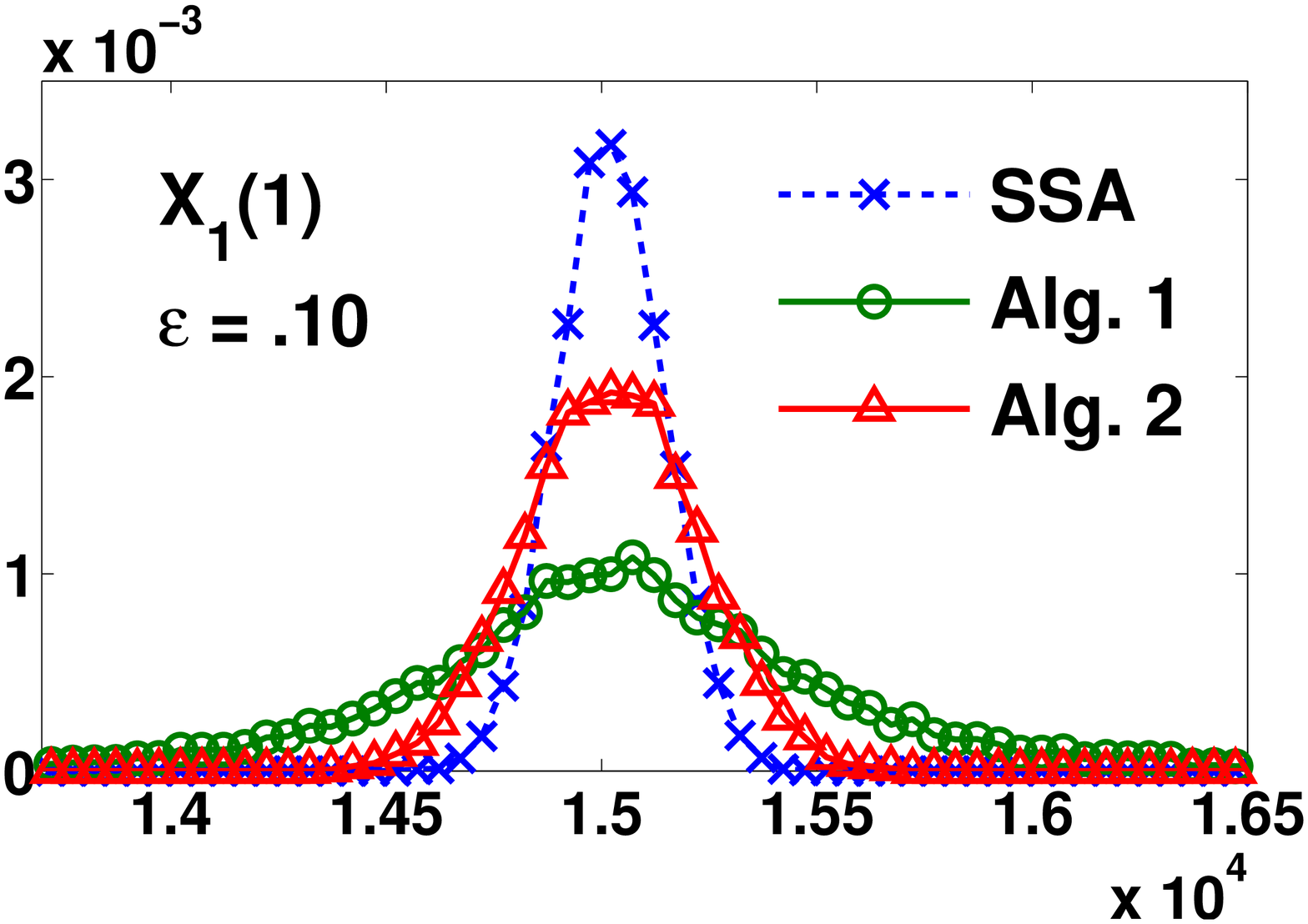}
    \includegraphics[width = 2.9in,height=2.2in]{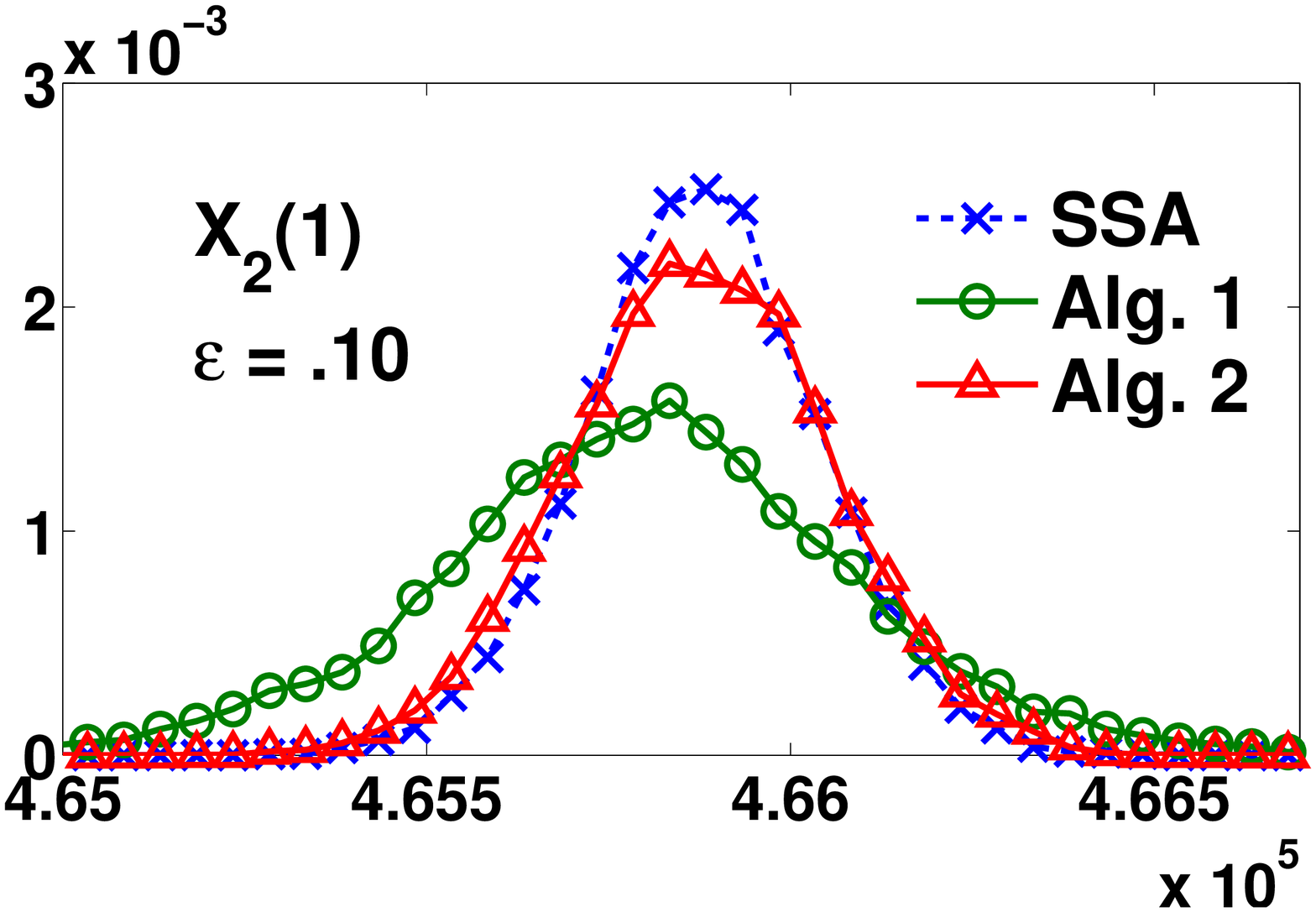}\\
    \includegraphics[width = 2.9in,height=2.2in]{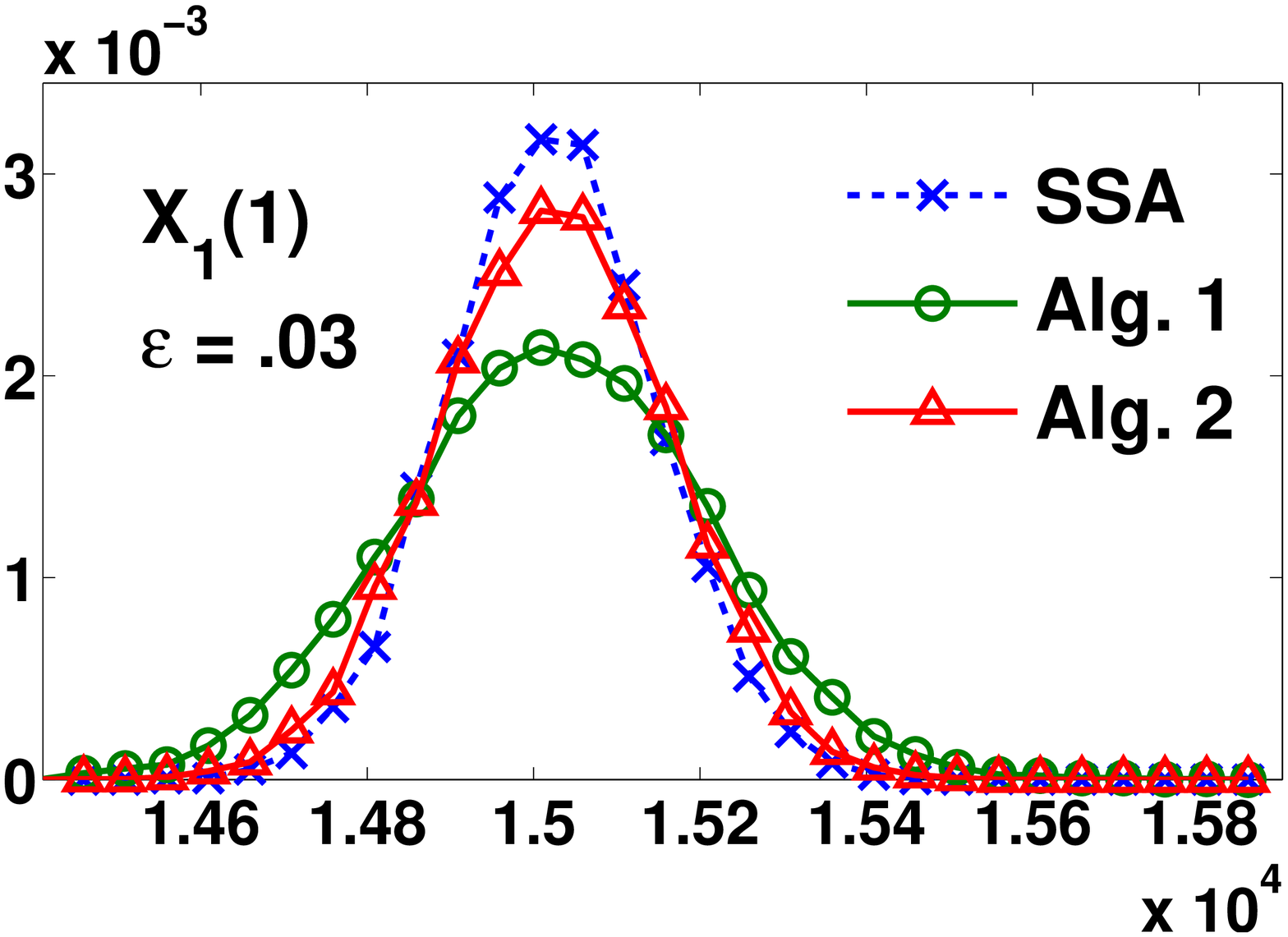}
    \includegraphics[width = 2.9in,height=2.2in]{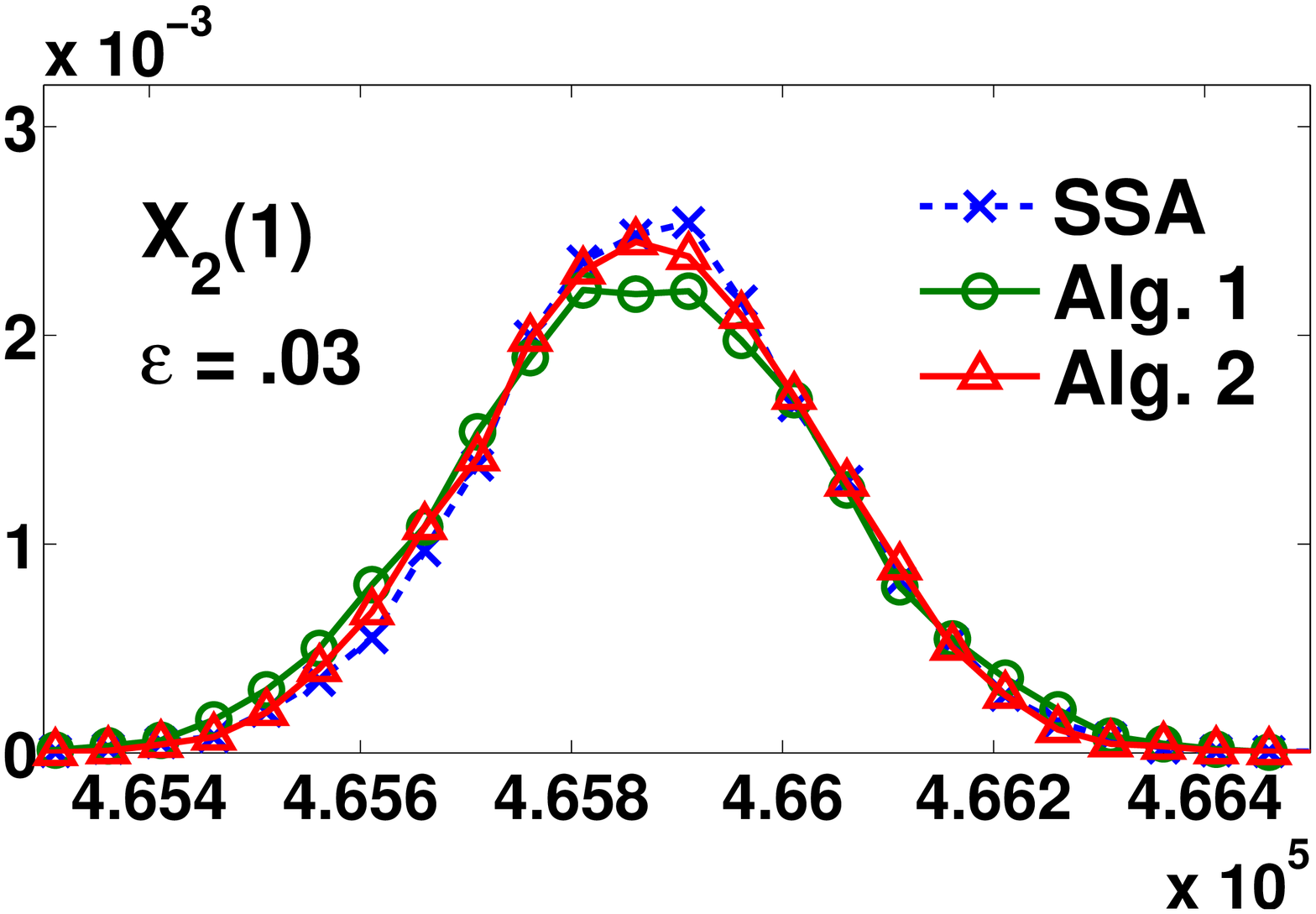}
  \end{center}

\pagebreak

Figure 2%\ref{fig:compare}

\vspace{1in}

\begin{center}
  \includegraphics[width = 2.9in,height=2.2in]{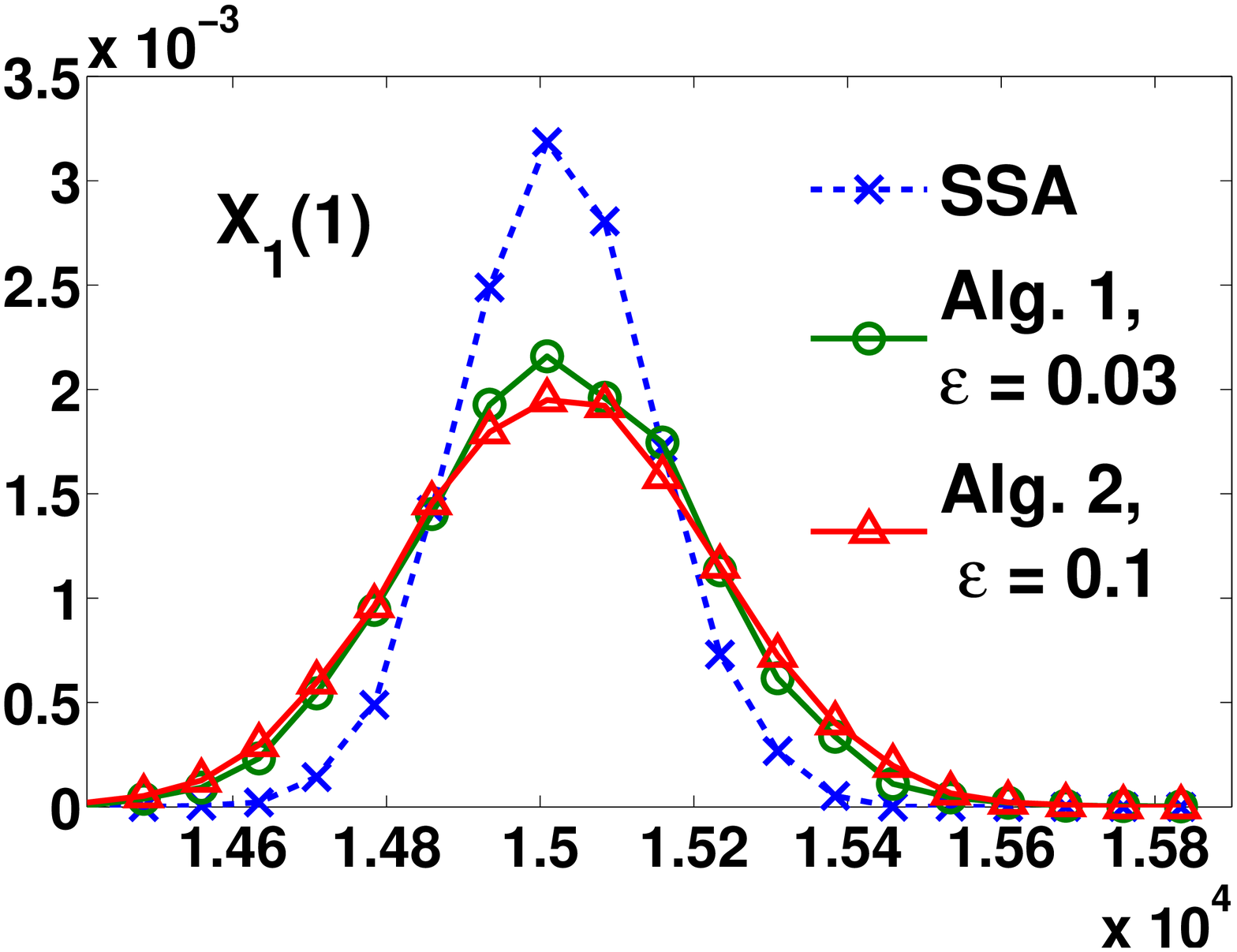}
  \includegraphics[width = 2.9in,height=2.2in]{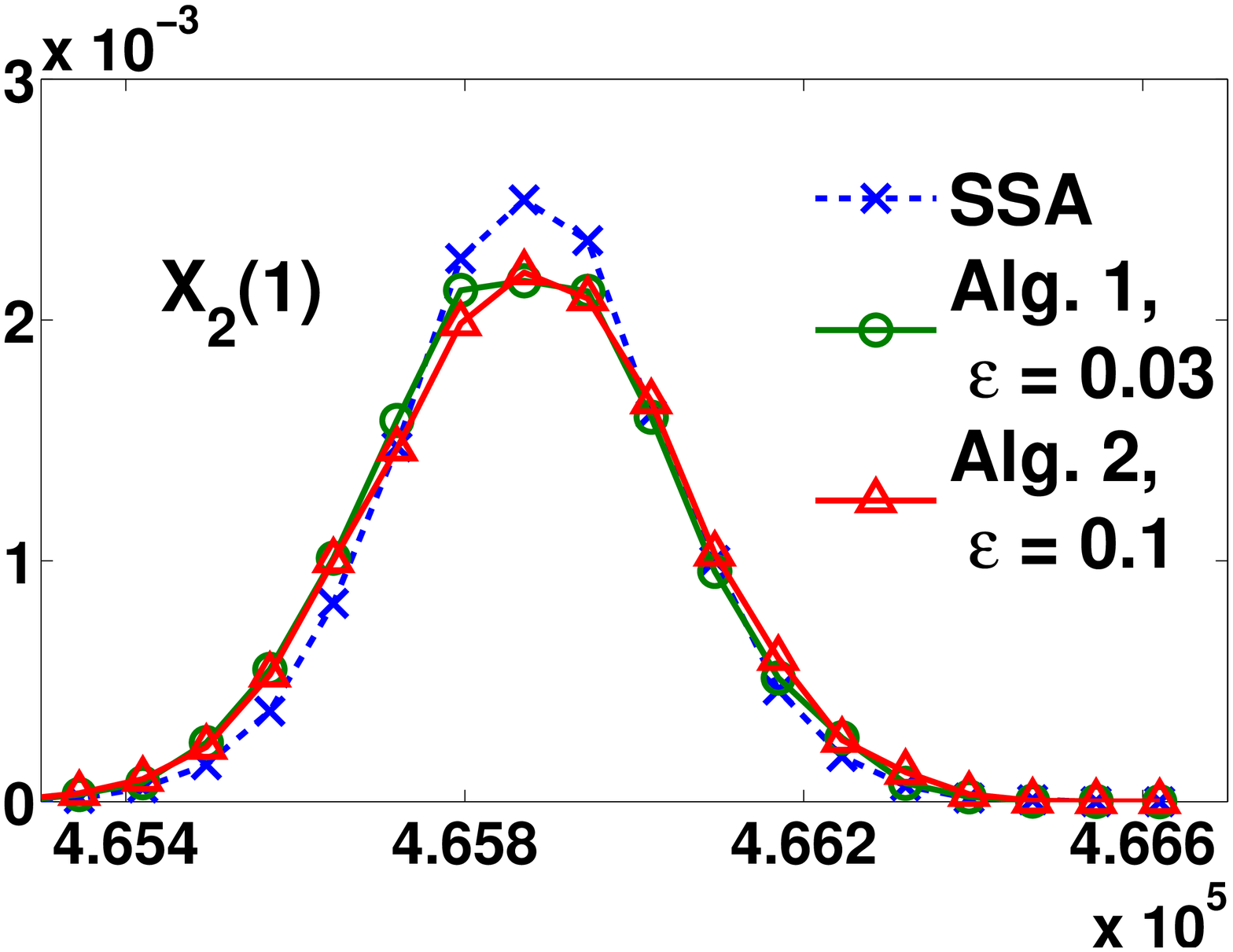}
\end{center}

\pagebreak

\noindent Captions

\vspace{1in}

\noindent Caption for Figure 1.

\vspace{.2in}

Histogram plots of $X_1(1)$ and $X_2(1)$ for $\epsilon = 0.03, 0.10,
0.17$.  Each plot was generated by simulating the system \eqref{dimer}
$10^4$ times using the SSA (dashed curve with `x'), Algorithm 1 (solid
curve with `o'), and Algorithm \ref{alg:fulltau} (solid curve with
`$\triangle$').  Algorithm \ref{alg:fulltau} is consistently more
accurate than Algorithm \ref{alg:oldtau}, as was expected.

\vspace{1in}

\noindent Caption for Figure 2.

\vspace{.2in}

Histogram plots of $X_1(1)$ and $X_2(1)$ found from $10^4$ simulations
of system \eqref{dimer}.  We show the SSA (dashed curve with `x'),
Algorithm \ref{alg:oldtau} with $\epsilon = 0.03$ (solid curve with
`o'), and Algorithm \ref{alg:fulltau} with $\epsilon = 0.1$ (solid
curve with `$\triangle$').  The distributions of the sample paths
generated by Algorithms \ref{alg:oldtau} and \ref{alg:fulltau} with
these $\epsilon$ values have similar accuracies and can therefore be
used as a fair test for efficiency.

\vspace{1in}

\noindent Caption for Table on page 17.

\vspace{.2in}

CPU times needed for Algorithms 1 and 2 to complete $10^4$ simulations
of system \eqref{dimer} for different $\epsilon$ values.

\end{document}